\documentclass[twocolumn,twoside,slac_two]{revtex4}
\usepackage{epsfig}
\usepackage{graphicx}
\newcommand{\PO}{I\!\!P}

\newcommand{\xpom}{x_{\PO}}

\begin{document}

\title{Review of diffraction at HERA and Tevatron}

%

\author{Laurent Schoeffel}
\affiliation{CEA Saclay, DAPNIA-SPP, 91191 Gif-sur-Yvette Cedex, France}
%

\begin{abstract}
We  present and discuss the  recent results on
diffraction from the HERA and Tevatron experiments.
\end{abstract}

\maketitle


\section{Diffraction at HERA}

\begin{figure}[!]
\begin{center}
\psfig{figure=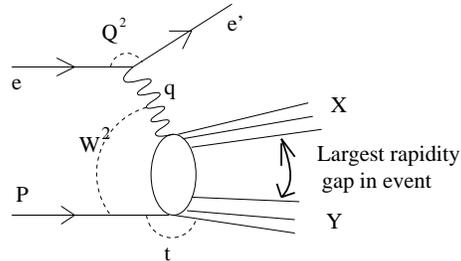,width=0.35\textwidth,angle=0}
\end{center}
\vspace{-0.5cm}
\caption{Illustration of the process $ep \rightarrow eXY$. The
hadronic final state is composed of two distinct systems $X$ and
$Y$, which are separated by the largest interval 
in rapidity between final state hadrons.}
\label{difproc}
\end{figure}

\begin{figure}[!]
\centering \epsfig{file=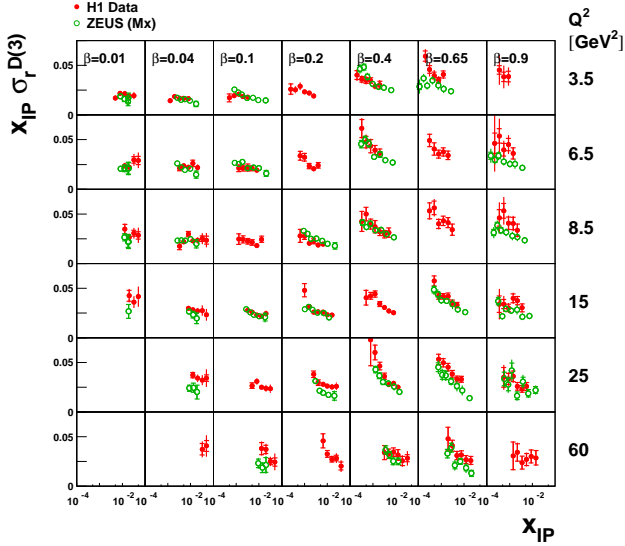,width=1.\linewidth}
\caption{Diffractive structure function measurements of the H1 and ZEUS
collaborations at HERA. The observable $\sigma_r^{D(3)} \simeq F_2^{D(3)}$ is determined from the
measured cross section using the relation : 
$
\frac{d^3 \sigma^{ep\rightarrow eXp}}{d\xpom\ dx\ 
dQ^2}=\frac{4\pi\alpha_{em}^2}{xQ^4}
({1-y+\frac{y^2}{2}})\sigma_r^{D(3)}(\xpom,x,Q^2)
$.
}
\label{fig1}
\end{figure}

At low $x$ in deep inelastic scattering (DIS) at HERA, approximately $10$ \% of the events
are of the type $ep \rightarrow eXp$, where the final state proton
carries more than $95$ \% of the proton beam energy \cite{royon}. For these processes,
a photon of virtuality $Q^2$, coupled to the electron (or positron),
undergoes a strong interaction with the proton (or one of its 
low-mass excited states $Y$) to form a hadronic final state
system $X$ of mass $M_X$ separated by a large rapidity gap (LRG)
from the leading proton (see Fig. \ref{difproc}). These events are called diffractive.
In such a reaction, $ep \rightarrow eXp$,
no net quantum number is exchanged and 
  the longitudinal momentum fraction $1-x_{\PO}$  
  is lost by the proton. Thus, the mongitudinal momentum $\xpom P$ is transfered 
to the system $X$. In addition, 
the virtual photon couples to a quark carrying
a fraction $\beta={x_{Bj}}/{x_{\PO}}$ of the exchanged momentum,
where $x_{Bj}$ is the Bjorken variable.
Extensive measurements of diffractive DIS cross sections have been made by both
the ZEUS and H1 collaborations at HERA,
using different experimental techniques \cite{data}. The results
are presented in Fig.~\ref{fig1}.

Events with the diffractive topology can be analysed in Regge models in terms of
Pomeron trajectory exchanged between the proton and the virtual photon.
Then, these events result from a colour-singlet exchange
between the diffractively dissociated virtual photon and the proton. 
Since this phenomenon is present even for high
virtuality of the photon,
it is called hard diffraction, at variance with soft diffractive phenomena.
Several theoretical formulations have been proposed.

\begin{figure}[!]
\centerline{\includegraphics[width=0.9\columnwidth]{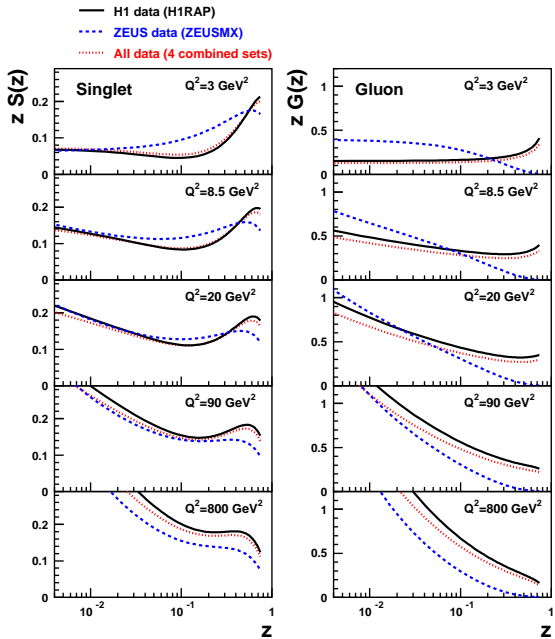}}
\caption{Singlet and gluon distributions 
of the Pomeron as a function of $z \equiv \beta$, the fractional momentum of the
Pomeron carried by the struck parton (see text) \cite{lolo}. }
\label{gluon}
\end{figure}

Among the most popular models, the one based on a pointlike structure of
the Pomeron has been studied quantitatively 
using a non-perturbative input supplemented by a perturbative QCD evolution equations \cite{lolo}.
In this formulation, it is assumed that the exchanged object, the Pomeron, 
is a colour-singlet quasi-particle whose structure is probed in
DIS process. 
As for standard DIS,   diffractive parton distributions (PDFs)  
related to the Pomeron can be derived from QCD fits to diffractive cross sections \cite{lolo}.
Diffractive PDFs extracted from H1 and ZEUS data are shown in Fig. \ref{gluon}. We
notice the large uncertainty (of about 25 \%) at large values of $\beta$
using different data sets \cite{lolo}.

There exists a different approach in which the cross sections 
are determined by the interaction between colour dipole states 
describing the photon and the proton \cite{p2}.
Indeed, it is well-known that the photon can be analyzed in terms of $q\bar{q}$ 
configurations while
it has been shown that the small-$x_{Bj}$ structure functions of the proton
can be described by a collection of primordial dipoles with subsequent perturbative
QCD evolution. This approach leads to a unified description of the proton total and
diffractive structure functions and provides a natural explanation for the 
geometric scaling property observed on the data \cite{golec2,plb}. In Fig. 4, diffractive 
cross sections,
$\beta\ \frac{d\sigma^{\gamma^*p\rightarrow Xp}_{diff}}{d\beta}
=\frac{4\pi^2\alpha_{em}}{Q^2} \xpom F_2^{D(3)}
$,
are presented as a fonction of the scaling variable $\tau_d = Q^2/Q_s^2$, where
$Q_s^2$ is the saturation scale combining variables $Q^2$, $x$ and $\xpom$
as defined in Ref. \cite{plb}. For each values of $\beta$, 
the function $\beta\ \frac{d\sigma^{\gamma^*p\rightarrow Xp}_{diff}}{d\beta}(\tau_d)$
is then observed to be continuous : it defines what is called the
geometrical scaling behaviour in this context \cite{plb}.


\begin{figure}[ht]
\begin{center}
\epsfig{file=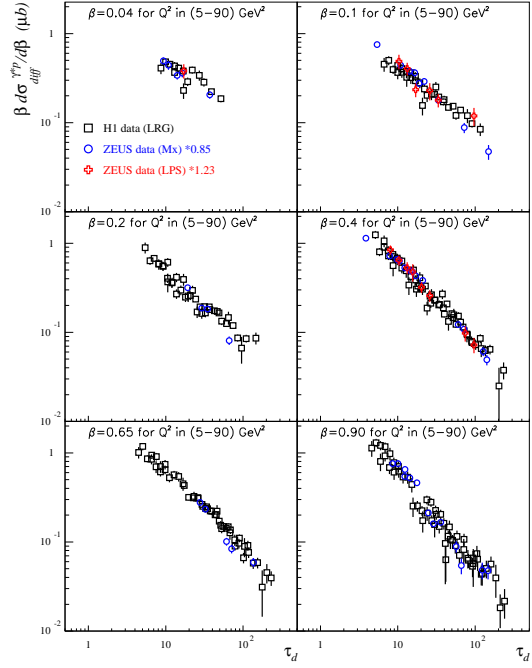,width=7cm}
\caption{The diffractive cross section
$\beta\ d\sigma^{\gamma^*p\rightarrow Xp}_{diff}/d\beta$ from H1 and ZEUS 
measurements, as a function of $\tau_d$ in bins of $\beta$ for $Q^2$ values in 
the range $[5;90]\ \mbox{GeV}^2$ and for $\xpom\!<\!0.01$ 
(see text) \cite{plb}.}
\end{center}
\label{fig2b}
\end{figure}


\section{Diffraction at Tevatron and LHC}

The difference between diffraction at HERA and at the Tevatron is that
diffraction at the Tevatron can occur not only on either $p$ or $\bar{p}$ side as at 
HERA, but also on both sides. The former case is called single diffraction
and the other one, double Pomeron exchange. In the same way
as we have defined the kinematical variables $\xpom$ and $\beta$ at HERA, we define $\xi_{1,2}$
as the protons fractional momenta loses  and $\beta_{1,2}$, the fractions of the
Pomeron momentum carried by the interacting partons. The produced diffractive
mass is equal to $M^2= s \xi_1 $ for single diffractive events and to
$M^2= s \xi_1 \xi_2$ for double Pomeron exchange,
where $\sqrt{s}$ is the energy of the reaction in the center of mass frame. The size of the rapidity gap
is then of the order of $\Delta \eta \sim \log 1/ \xi_{1,2}$ \cite{royon,ls1}.

\begin{figure}[!]
\begin{center}
\epsfig{file=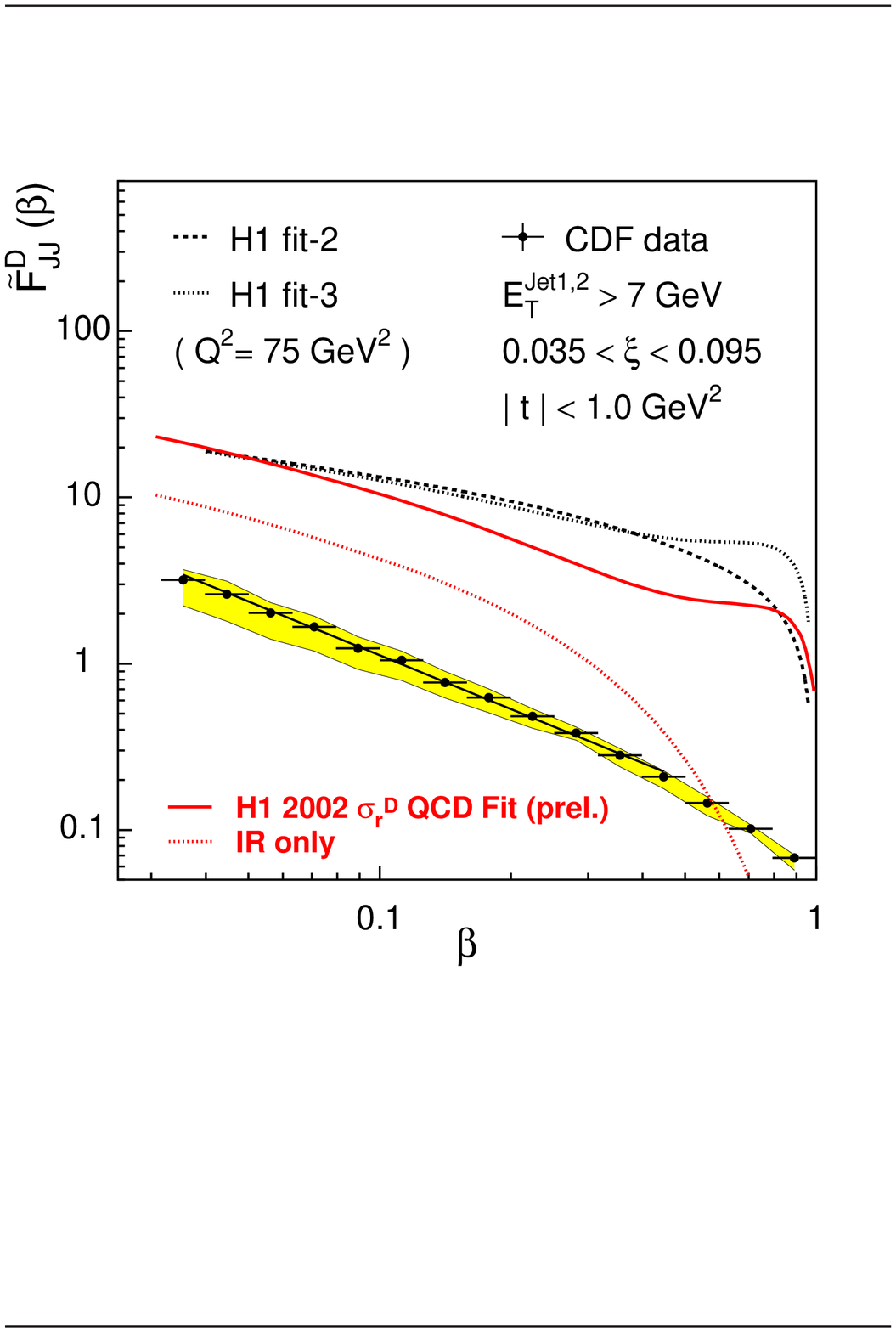,width=7cm,clip=true}
\vspace{6.cm}
\caption{Comparison between the CDF measurement of diffractive structure
function (black points) with the expectation of the HERA diffractive PDFs.}
\label{cdfh}
\end{center}
\end{figure}

It has been shown that the diffractive PDFs of HERA can not be used directly to make predictions
at the Tevatron. Indeed, factorisation does not hold and a gap survival probability of a few \% has to be considered
(see Fig. \ref{cdfh}).
It corresponds to the
probability that there is no soft additional interaction or in other words that
the event remains diffractive \cite{royon,ls1}. 

\begin{figure}[!]
\begin{center}
\includegraphics[totalheight=7cm]{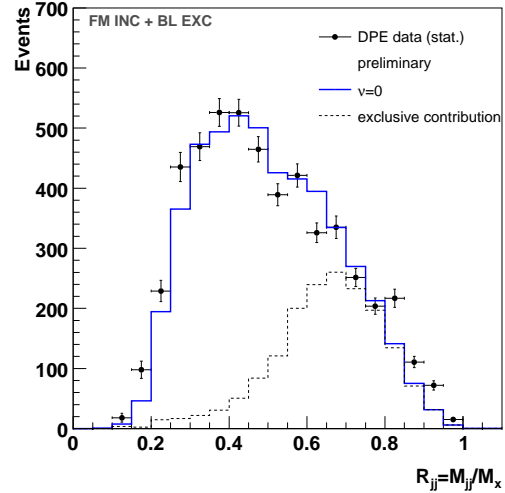}
\caption{Dijet mass fraction for jets $p_T>10\,\mathrm{GeV}$. 
The data are compared to the sum of inclusive and
exclusive predictions. The dPDFs derived from H1 data have been used together
with the survival gap probability measured with single diffractive events at Tevatron.}
\label{dijetmass}
\end{center}
\end{figure}

The CDF collaboration has measured the so-called dijet mass fraction (DMF) in dijet
events when the antiproton is tagged in the roman pot detectors
 and when there is a rapidity gap on the proton side to ensure that the
event corresponds to a double Pomeron exchange. 
The measured observable  $R_{jj}$ is defined as the ratio of the mass carried by the two jets divided by the total
diffractive mass \cite{royon, ls1, kepka}. 
The DMF turns out to be a very appropriate observable for identifying the exclusive 
production, which would manifest itself as an excess of the 
events towards $R_{jj}\sim 1$ \cite{royon, ls1, kepka}. 
Indeed, for exclusive events, the dijet mass is essentially equal
to the mass of the central system because no Pomeron remnant is present.
Then, for exclusive events, the
DMF is 1 at generator level and can be smeared
out towards lower values taking into account the detector resolutions.
The advantage of the DMF is that one can focus on the shape of the distribution
and let the absolute normalisation of the predictions as free parameters (in a first step). Indeed, the observation 
of exclusive events does not rely on the overall normalization which might be strongly dependent on
the detector simulation and acceptance of the roman pot detector.
Results are shown
in Fig~\ref{dijetmass} with Monte-Carlo (MC) expectations. These MC predictions are displayed for 
inclusive double Pomeron exchange events, where remnants are still present in the reaction
and then $R_{jj}$ is lower that unity, as well as for exclusive events in a specific model.
More details are given on the models entering into these simulations in Ref. \cite{royon,  kepka}. 
We observe a 
good agreement when   both contributions, inclusive and exclusive, are taken into account.
A MC with inclusive simulation only is not sufficient to describe the present data, and shows a
clear deficit of events towards high values of the DMF, 
where exclusive events are supposed to occur. 
This part of the DMF is properly described by the exclusive contribution, once normalised to fit the histogram.
It is a first evidence that exclusive events could contribute at the Tevatron \cite{royon,kepka}.

The great interest of studying such exclusive dijet events is that it opens the possibility to
analyse the production of heavy objects in double Pomeron exchange at the LHC \cite{royon, ls1, kepka}.
In particular, the production of a Higgs boson in such a topology could be quite interesting as
the event would be very clean : both protons escape and are detected in Roman pots, two large rapidity gaps
on both sides and the central production of the Higgs boson, leading to some decay products well
isolated in the detector  \cite{royon, ls1, kepka}.
The major advantage of such events is that the resolution on the mass of
the  produced object can be determined with a high resolution from the
measurement of the proton momentum loses, using the relation $M^2= s \xi_1 \xi_2$.
A potential signal, accessible in a mass distribution, is then not washed out by the lower resolution
when using central detectors, rather than forward Roman pots to measure $\xi_1$ and $\xi_2$.

\section{Conclusions}

We have presented and discussed the most recent results on
inclusive diffraction from the HERA and Tevatron experiments.
Of special interest for future prospects is the exclusive
production  heavy objects (including Higgs boson) which requires a good understanding of diffractive processes 
and the link between electron-proton
and hadronic colliders.


\end{document}